\documentclass[journal]{IEEEtran}
\usepackage{amsmath,amssymb}
\usepackage{graphicx}
\graphicspath{{figures/}}
\usepackage{booktabs}
\usepackage{algorithm}
\usepackage{algpseudocode}
\usepackage{cite}
\usepackage{url}
\usepackage{xcolor}
\newcommand{\todo}[1]{\textcolor{red}{[#1]}}

\begin{document}
\title{Privacy-preserving federated tensor decomposition of single-cell immune data:\\
recovering multicellular programs across institutions}
\author{Axel Faes,
  St\'ephanie M.~van den Berg, and Maryam Amir Haeri%
  \thanks{A. Faes is with the Data Science Institute, Hasselt University,
  Belgium, and with the University of Twente, The Netherlands
  (e-mail: axel.faes@utwente.nl).}%
  \thanks{S. M. van den Berg and M. Amir Haeri are with Learning, Data
  Analytics and Technology, University of Twente, The Netherlands.}}

\maketitle

\begin{abstract}
Tensor decomposition of donor $\times$ cell-type $\times$ gene single-cell data recovers
\emph{multicellular programs}: coordinated axes of inter-individual transcriptional variation that
span cell types and stratify disease. Such analyses assume all cells are centralised, yet immune
single-cell atlases are increasingly multi-institution, multi-ancestry, and governed: patient cells
often cannot be pooled. We present a federated estimator for multicellular tensor decomposition.
Each site computes a low-dimensional local program subspace; a coordinator merges these by stacked
singular value decomposition under federated global-mean centering, an operation provably equivalent
(up to truncation) to the centralised decomposition. This centering also makes the merge robust to
site-label confounding (program AUC held flat at $0.957$ across case-fraction skew, whereas naive
per-site centering degrades to $0.861$). Only program subspaces (never cells or
donor-level scores) leave a site, and aggregation is a summation compatible with secure aggregation.
On a 261-donor systemic lupus erythematosus (SLE) atlas the federated estimator recovers the canonical
interferon multicellular program (gene-mode ISG enrichment AUC $0.998$; case--control separation
$0.958$; bootstrap $\Delta\text{AUC}=-0.000$, 95\% CI $[-0.004,+0.012]$ vs.\ centralised; label-free
cross-validated AUC $0.960$), faithfully across institution-scale and multi-ancestry partitions, and
across the three \emph{real} collection sites of a COVID-19 atlas (subspace correlation $0.989$). We
further show the estimator recovers the full multicellular program when \emph{no site observes all
cell types} (subspace correlation $1.000$, exact by construction), a setting fixed-feature federated
PCA cannot address; it is private at the complete-panel level when the donor coupling is computed under
secure computation and only loadings are released. On an interstitial-lung-disease atlas, where the
fibrotic response is intrinsically multicellular, the recovered program predicts disease better than the
best single cell type (AUC $0.96$ vs.\ $0.91$; gap $95\%$ CI excludes zero, robust across seeds) and the
advantage survives federation, the regime the method targets; a smaller fibrosing-cholangiopathy liver
cohort is consistent (label-specific by permutation, $p=0.005$); on
compartment-dominated blood, gut, kidney and heart cohorts it matches but does not exceed the best
single compartment. A
membership-inference evaluation quantifies the privacy gain of secure aggregation (attack AUC reduced
from $0.91$ to $0.61$ for the complete-panel scheme), directly addressing the donor re-identification
risk recently demonstrated for single-cell count matrices. The method enables cross-institution
and cross-ancestry recovery and comparison of multicellular immune programs without sharing cells.
\end{abstract}


\section{Introduction}
\IEEEPARstart{P}{opulation-scale} single-cell RNA sequencing has made it possible to study how the
\emph{coordinated} state of the immune system varies between individuals. An established framing
represents a patient cohort as a third-order tensor of donor $\times$ cell-type $\times$ gene
pseudobulk expression and decomposes it to recover \emph{multicellular programs}: factors whose
loadings span several cell types and capture inter-individual axes of variation that no single cell
type expresses in isolation. Interpretable tensor decomposition (scITD) introduced this view and used
it to stratify lupus and COVID-19 patients~\cite{scitd}; related tensor methods model cell--cell
communication across contexts~\cite{tensorcell2cell} and condition-resolved variation~\cite{parafac2rise}.
These methods are \emph{centralised}: they assume every cell is available in one place.

Such programs are of direct biomedical interest: they place each patient on continuous axes of immune
state, stratify cohorts beyond case/control labels, and link coordinated transcriptional shifts to
disease severity, treatment response, and genetics. Their value grows with cohort size and diversity,
so the most informative analyses are precisely those spanning many institutions and ancestries, the
setting hardest to centralise.

In biomedical practice that assumption rarely holds. Immune single-cell atlases are assembled across
hospitals, biobanks, and national cohorts~\cite{regev2017human}, and the underlying data are governed: patient-derived
single-cell profiles are subject to consent and data-protection constraints that frequently preclude
pooling raw cells~\cite{rieke2020future,sheller2020federated}. Multi-ancestry comparisons compound the problem, since the most informative
contrasts often span populations held by different consortia. Federated analysis (computing over data
that remain at their source) is the natural response~\cite{mcmahan2017fedavg,kairouz2021advances}, and federated methods now exist for single-cell
batch correction~\cite{fedscgen}, cell-type classification under homomorphic encryption~\cite{pricell},
and quantitative-trait-locus mapping~\cite{privateqtl}. None recovers multicellular \emph{programs}:
the tensor-decomposition task that scITD addresses centrally has no federated counterpart.

We close that gap. Our contributions are deliberately scoped and reported with their limits:
\begin{enumerate}
\item \textbf{Vertical-FL recovery of cross-panel multicellular programs}: recovery of the full
multicellular program when \emph{no site observes all cell types}, by coupling through the shared donor
mode. This is a vertical federated setting (shared donors, partitioned cell-type panels); it recovers
programs spanning cell types no single site sees, which fixed-feature federated PCA cannot handle at all.
\item A \textbf{confounding-robust federated estimator}: sites share only a low-dimensional program
subspace; the coordinator merges local subspaces by stacked SVD under federated global-mean centering,
reproducing the centralised decomposition up to truncation, and the aggregation is a summation hence
compatible with secure aggregation~\cite{secagg}. This complete-panel merge instantiates distributed
PCA~\cite{distpca,grammenos2020federated}; the methodological additions are that the global-mean centering makes the merge
robust to site-label confounding (Table~\ref{tab:centering}) and, with contribution~1, the cross-panel
donor coupling.
\item A \textbf{membership-inference privacy evaluation}~\cite{shokri2017membership} quantifying the leakage of each release
granularity and the protection conferred by secure aggregation, the practical guarantee on which our
privacy claims rest (formal differential privacy~\cite{dwork2014algorithmic} is out of reach at present single-cell cohort sizes;
Section~\ref{sec:disc}).
\item \textbf{A cross-tissue map of when multicellular recovery matters} (\S\ref{sec:whenmatters}): across
seven cohorts the multicellular program out-predicts the best single cell type only where cross-cell-type
coordination is strong, with interstitial lung disease the stated result (cross-validated AUC $0.959$ vs.\
$0.910$, gap $95\%$ CI excluding zero); a smaller fibrosing-cholangiopathy liver cohort corroborates
(separation label-specific by permutation, $p=0.005$), and the program is at parity on
compartment-dominated blood, gut, kidney and heart, establishing when the method earns its cost.
\end{enumerate}
We are explicit about scope (Section~\ref{sec:disc}): multicellular decomposition out-predicts the best
single cell type only where cross-cell-type coordination is strong (interstitial lung disease and
fibrosing-cholangiopathy liver) and is at
parity on compartment-dominated blood, gut, kidney and heart; the central contribution is faithful
\emph{private federated recovery} of established coordinated programs, and formal differential privacy is
left to future work.

\section{Related Work}
\textbf{Centralised multicellular decomposition.} scITD~\cite{scitd} factorises a donor $\times$
cell-type $\times$ gene tensor by Tucker decomposition to recover multicellular programs;
Tensor-cell2cell~\cite{tensorcell2cell} models cell--cell communication across contexts and
PARAFAC2-RISE~\cite{parafac2rise} condition-resolved variation, while MOFA~\cite{mofa} integrates
modalities by matrix factorisation. DIALOGUE~\cite{jerbyarnon2022dialogue} recovers multicellular
programs by cross-cell-type association rather than tensor decomposition. All operate on pooled data.

\textbf{Federated single-cell analysis.} Existing federated single-cell methods target different
tasks: FedscGen~\cite{fedscgen} performs batch-effect correction, PriCell~\cite{pricell} cell-type
classification under homomorphic encryption, and foundation-model pretraining has been federated as
well~\cite{tabula}. None recovers multicellular \emph{programs}; the tensor-decomposition task has no
federated counterpart, which is the gap we address.

\textbf{Federated and private factorisation.} Distributed PCA over partitioned data is well
characterised~\cite{distpca,grammenos2020federated}, and the donor-mode SVD we federate is an instance. Our methodological
contribution is therefore not the complete-panel merge per se but (i) the vertical-FL donor coupling that
recovers cross-panel programs no fixed-feature method addresses, and (ii) the global-mean centering that
makes the merge robust to site-label confounding. Differentially
private low-rank factorisation has been studied for PCA and tensor
decompositions~\cite{noisypower,dppca2022,dwork2014analyzegauss,chaudhuri2013near}, but statistically meaningful private subspace recovery needs
$n=\widetilde O(d)$ samples~\cite{dppca2022}, beyond current single-cell atlases (Section~\ref{sec:disc}).
Secure aggregation~\cite{secagg} provides the lossless aggregation primitive our complete-panel scheme
uses, and our privacy claims rest on it together with the membership-inference evaluation below.

\textbf{Federated genomics privacy.} privateQTL~\cite{privateqtl} performs secure federated
quantitative-trait-locus mapping via multiparty computation, illustrating the governance pressure that
motivates federation of patient molecular data. The risk is concrete: donor re-identification has been
demonstrated directly from single-cell count matrices~\cite{homer2008resolving,erlich2014routes,walker2024leakage}, with differential privacy named as a
still-unbuilt defence. Our contribution is complementary: we federate the
\emph{unsupervised multicellular program} rather than QTL association, and we characterise the privacy
of the resulting release objects under secure aggregation.

\section{Methods}

\subsection{Multicellular tensor and centralised decomposition}
For $D$ donors, $C$ cell types, and $G$ genes, let $r_{d c g}$ denote the summed UMI count of gene $g$
over all cells of type $c$ in donor $d$ (pseudobulk)~\cite{crowell2020muscat,squair2021confronting}. We normalise each (donor, cell type) profile to
$\log$ counts-per-million, $\tilde r_{d c g}=\log\!\big(1+10^{6}\,r_{d c g}/\sum_{g'} r_{d c g'}\big)$,
retain the $G$ most variable genes, and \emph{standardise each gene within each cell type across
donors}:
\begin{equation}
\mathcal{T}_{d c g}=\frac{\tilde r_{d c g}-\mu_{c g}}{\sigma_{c g}},\qquad
\mu_{c g}=\tfrac{1}{|\mathcal D_c|}\!\sum_{d\in\mathcal D_c}\!\tilde r_{d c g},
\label{eq:norm}
\end{equation}
with $\sigma_{c g}$ the corresponding standard deviation and $\mathcal D_c$ the donors in which type
$c$ is observed ($\geq 20$ cells); slabs absent in a donor are masked. This centring removes cell-type
baseline so $\mathcal{T}\in\mathbb{R}^{D\times C\times G}$ carries \emph{inter-individual} variation,
the scITD convention~\cite{scitd}. The donor-mode unfolding $X\in\mathbb{R}^{D\times CG}$ flattens the
cell-type and gene axes~\cite{kolda2009tensor,delathauwer2000multilinear}. A rank-$K$ decomposition takes the top-$K$ right singular vectors
$V\in\mathbb{R}^{CG\times K}$ of the column-centred $X-\bar x$; each column, reshaped to $C\times G$, is
a \emph{multicellular program} whose per-cell-type loading norms indicate which cell types participate,
and the donor scores $U=(X-\bar x)V\in\mathbb{R}^{D\times K}$ place each donor on every program. We use
$K=10$ and $G=1500$ throughout; donors are retained if at least four cell types are present.

\subsection{Federated estimator}
\label{sec:fed}
Donors are partitioned across $S$ sites; site $s$ holds rows $X_s$ and never transmits them. The
coordinator obtains the federated global mean $\bar x=\tfrac1D\sum_s n_s\bar x_s$ from per-site means.
Each site centres locally, computes a truncated SVD $X_s-\bar x=U_s\Sigma_s V_s^\top$, and transmits
only the scaled loadings $M_s=\Sigma_s^{(K)}V_s^{(K)\top}\in\mathbb{R}^{K\times CG}$. The coordinator
stacks $M=[M_1;\dots;M_S]$ and takes the top-$K$ right singular vectors of $M$ as the global program
basis $V_{\mathrm{fed}}$; donor scores are computed locally by projection. Since
\begin{equation}
M^\top M=\textstyle\sum_s V_s\Sigma_s^2 V_s^\top=\sum_s (X_s-\bar x)^\top(X_s-\bar x)=(X-\bar x)^\top(X-\bar x),
\end{equation}
the stacked-SVD basis spans the same subspace as the centralised decomposition up to truncation
error~\cite{distpca}; the merge also resolves the per-site rotation gauge, so no explicit factor
alignment is required (Algorithm~\ref{alg:complete}).

\begin{algorithm}[t]
\caption{Federated complete-panel multicellular decomposition}
\label{alg:complete}
\begin{algorithmic}[1]
\State \textbf{Input:} site unfoldings $\{X_s\}_{s=1}^S$, rank $K$
\State each site sends $(n_s,\bar x_s)$; coordinator sets $\bar x=\tfrac1D\sum_s n_s\bar x_s$, broadcasts
\For{each site $s$ in parallel}
  \State $U_s\Sigma_s V_s^\top \gets \mathrm{SVD}(X_s-\bar x)$;\; send $M_s=\Sigma_s^{(K)}V_s^{(K)\top}$
\EndFor
\State coordinator (secure aggregation): $V_{\mathrm{fed}}\gets$ top-$K$ right vectors of $[M_1;\dots;M_S]$
\State each site computes donor scores $U_s=(X_s-\bar x)V_{\mathrm{fed}}$ \emph{locally}
\State \textbf{Output:} program basis $V_{\mathrm{fed}}$
\end{algorithmic}
\end{algorithm}

\subsection{Complexity and communication}
Each site computes one truncated SVD, $O(n_s\,CG\,K)$, and transmits $K\times CG$ floats per round
\emph{independent of its cell count}; the coordinator's merge is $O(SK\,CG\,K)$. Pooling, by contrast,
would transfer the $\sim\!10^6$ raw cells $\times CG$ entries per site. The complete-panel scheme is
single-round (the stacked-SVD merge is exact up to truncation, so no iterative communication is needed).

\subsection{Cell-type-incomplete federation}
When sites profile \emph{different} cell-type panels, the (cell-type$\times$gene) feature space differs
per site and fixed-feature federated PCA does not apply. This is a \emph{vertical} federated setting
(shared donors, partitioned features)~\cite{yang2019federated}. We exploit the shared donor mode: with donors common across
sites and disjoint cell-type panels, each site contributes a donor-gram block
$G_s=(X_s-\bar x)(X_s-\bar x)^\top$ over \emph{its} cell-type columns; the coordinator sums
$G=\sum_s G_s=(X-\bar x)(X-\bar x)^\top$ and eigendecomposes it to the shared donor scores, after which
each site returns the loadings for its own cell types and the coordinator concatenates them into the
full $C\times G$ program, spanning cell types no single site observed. The summed Gram \emph{equals}
the centralised Gram, so recovery is exact \emph{by construction}: the contribution is the
\textbf{capability} (fixed-feature federated PCA cannot recover cross-panel programs at all), not an
empirical accuracy gain. This capability carries a privacy cost: the donor Gram exposes donor-level
structure, which we quantify in the privacy evaluation below (it requires DP or homomorphic
encryption on the Gram~\cite{gentry2009fully,cheon2017homomorphic,mohassel2017secureml}, unlike the complete-panel subspace scheme); cross-site recovery also presumes
donors are matched across sites (a private-set-intersection step) and harmonised cell-type annotations
(Algorithm~\ref{alg:incomplete}).

\begin{algorithm}[t]
\caption{Incomplete-panel recovery (vertical FL, donor-mode coupling)}
\label{alg:incomplete}
\begin{algorithmic}[1]
\State \textbf{Input:} shared donors; site cell-type panels $\{C_s\}$ (disjoint, $\cup_s C_s=$ all types)
\For{each site $s$ in parallel}
  \State $G_s\gets (X_s-\bar x)(X_s-\bar x)^\top$ over its columns $C_s$ \Comment{$D\times D$}
\EndFor
\State coordinator (HE/MPC): $G\gets\sum_s G_s$; \; $U\gets$ top-$K$ eigenvectors of $G$
\For{each site $s$ in parallel}
  \State $V_s\gets X_s^\top U$ for cell types $C_s$ \Comment{computed under HE/MPC; only loadings revealed}
\EndFor
\State coordinator concatenates $\{V_s\}$ into the full $C\times G$ program $V$
\State \textbf{Output:} full multicellular program $V$ (no site observed all $C$)
\end{algorithmic}
\end{algorithm}

\subsection{Privacy model and membership inference}
\textbf{Threat model.} We assume an honest-but-curious coordinator and honest-but-curious sites: all
follow the protocol but may inspect any message they receive. The only quantities leaving a site are
per-site means and scaled loadings $M_s$ (or Gram blocks $G_s$), population-level, low-dimensional
summaries; raw cells and donor-level scores $U$ never leave a site. Secure aggregation~\cite{secagg}
guarantees the coordinator observes only the aggregate $\sum_s M_s$ (resp.\ $\sum_s G_s$), not any
individual site's contribution.

\textbf{Membership inference.} Secure aggregation hides intermediate messages but does not by itself
bound what the \emph{released} program reveals about a donor. We quantify this with a
reconstruction-residual membership-inference attack~\cite{shokri2017membership,carlini2022membership}. For a released $CG\times K$ subspace $V$ fit on a
member set $\mathcal M$, define a donor's projection residual
$\rho(x)=\lVert(x-\bar x)-(x-\bar x)VV^\top\rVert_2$. Donors in $\mathcal M$ tend to have smaller
$\rho$ than held-out non-members; the attacker's membership score is $-\rho$, and the attack AUC over
$30$ balanced member/non-member splits measures leakage ($0.5=$ none). We evaluate four release
objects: (i)~per-donor scores $U$; (ii)~per-site subspaces $\{V_s\}$ (a curious coordinator
\emph{without} secure aggregation); (iii)~the merged subspace $V_{\mathrm{fed}}$ (\emph{with} secure
aggregation); and (iv)~the incomplete-panel donor Gram $G$, whose eigenvectors reconstruct $U$.

\subsection{Data and evaluation}
We use five public atlases. \textbf{SLE}: Perez et al.~\cite{perez2022}, 1.26\,M PBMCs from 261 donors
(162 SLE, 99 control), 11 cell types, Asian and European ancestry. \textbf{COVID-19}:
Stephenson et al.~\cite{stephenson2021}, PBMCs from 130 donors across three UK centres (Newcastle,
Cambridge, Sanger); we use the 117 donors with initial-timepoint samples, 18 compartments, and a
severity contrast over the 73 severity-graded donors (22 severe/critical/death vs.\ 51 asymptomatic/
mild/moderate). \textbf{IBD}: the IBDverse Crohn's-disease gut atlas~\cite{ibdverse} (used for the
cross-disease meta-analysis and the complementarity map), 9 compartments. \textbf{ILD}: the
interstitial-lung-disease atlas of Natri et al.~\cite{natri2024ild} ($109$ donors, $33$ cell types,
$63$ ILD vs.\ $46$ normal), the most strongly cross-cell-type-coordinated cohort. \textbf{Liver}: the
PSC/PBC fibrosing-cholangiopathy atlas of Andrews et al.~\cite{andrews2024psc} ($16$ donors, $10$ PSC/PBC
vs.\ $6$ normal, $27$ cell types; caudate-lobe scRNA), a second cross-cell-type-coordinated fibrotic
cohort. All are processed
identically (Eq.~\ref{eq:norm})~\cite{wolf2018scanpy,hao2021integrated}; cell types follow each study's published broad annotation, and the gut
and lung atlases use their respective tissues only.

Recovery is assessed by three metrics. \emph{(i) Subspace canonical correlation} between the centralised and
federated program bases $V_{\mathrm{c}},V_{\mathrm{f}}\in\mathbb{R}^{CG\times K}$: the mean of the
cosines of the principal angles, i.e.\ the mean singular value of $Q_{\mathrm c}^\top Q_{\mathrm f}$
for orthonormal bases $Q$ ($1.0$ = identical subspace). \emph{(ii) Program face validity}: a
Mann--Whitney test of whether $|$gene-mode loading$|$ is higher for a curated interferon-stimulated-gene
(ISG) set than for the remaining genes, reported as the corresponding AUC. \emph{(iii) Phenotype
separation}: the ROC AUC of the disease program's donor scores against the clinical label
(SLE case/control; COVID severe/critical/death vs.\ asymptomatic/mild/moderate), with donor-bootstrap
$95\%$ confidence intervals ($300$ resamples) on all AUCs, $\Delta$AUC, and subspace correlations. To
exclude test-set selection bias we additionally report a \emph{label-free} five-fold cross-validation
in which the disease program is chosen on each training fold by ISG enrichment alone (never by the
label) and scored on the held-out fold (per-fold AUC, sign-resolved).

\section{Results}

\subsection{Centralised recovery and biological validity}
On the SLE atlas the decomposition yields a dominant multicellular program: a single donor-mode factor
whose $C\times G$ loading spans all eleven cell types, peaking on classical monocytes but with coherent
positive loadings across B, T, NK, and dendritic compartments, i.e.\ a genuinely \emph{multicellular}
signature of coordinated inter-individual variation rather than a single-cell-type effect. Its donor
scores separate SLE from control at AUC $0.958$ (donor-bootstrap 95\% CI $[0.79,0.98]$; the absolute
value carries real uncertainty, whereas the federation-lossless result below is far better
constrained); its gene-mode loading is strongly enriched for ISGs (enrichment AUC
$0.998$, $p=7.4\times10^{-13}$) and its cell-type loading peaks on classical monocytes, the
established hallmark of the lupus type-I interferon response, matching centralised scITD~\cite{scitd}
and providing a ground-truth target.

\subsection{Federated recovery is faithful and lossless}
\begin{figure}[t]\centering
\includegraphics[width=\linewidth]{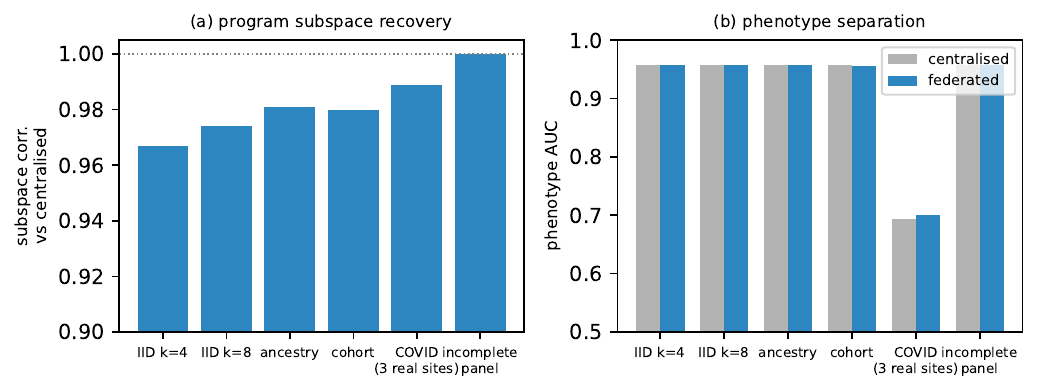}
\caption{Federated recovery is faithful across regimes. (a) Subspace correlation to the centralised
program ($\geq 0.96$, $1.000$ for the exact incomplete-panel scheme). (b) Phenotype-separation AUC,
federated vs.\ centralised (near-identical; COVID is the weaker severity program).}
\label{fig:recovery}
\end{figure}

The federated estimator reproduces the centralised program in every regime (Table~\ref{tab:rec},
Fig.~\ref{fig:recovery}).
Donor-bootstrap analysis gives $\Delta\text{AUC}=-0.000$ (95\% CI $[-0.004,+0.012]$; the interval
contains zero, i.e.\ federation is statistically lossless) and subspace correlation $0.967$
(CI $[0.92,0.99]$). Selecting the program \emph{without labels} (by ISG enrichment on training folds)
and evaluating on held-out donors gives AUC $0.960$, ruling out test-set selection bias. Recovery holds
under partitions that stress the estimator. A realistic \emph{multi-ancestry} split places
Asian-ancestry donors (78\% SLE) at one site and the rest (51\% SLE) at another, strong label skew of
the kind real consortia face, yet recovery is unchanged (AUC $0.958$, correlation $0.981$, ISG
enrichment $0.998$). An adversarial \emph{processing-cohort} split, in which one site contains $0\%$
cases and another $90\%$, is likewise recovered (AUC $0.956$, correlation $0.980$): because every site
centres on the federated global mean, the merge is insensitive to such site--label confounding.

On the COVID-19 atlas we test the estimator across the \emph{actual} three collection sites rather than
simulated partitions: Newcastle ($61$ donors), Cambridge ($45$), and the Wellcome Sanger Institute
($11$). Federation recovers the centralised severity-program subspace at correlation $0.989$ (severity
program cross-validated AUC $0.678$ over the $73$ severity-graded donors). The smallest site ($n{=}11$)
contributes a rank-deficient local subspace yet does not degrade the merge, since the stacked-SVD pools
rank across sites. This is, to our knowledge, the first multicellular-program recovery federated across
genuine institutional boundaries.

\subsection{Recovery from cell-type-incomplete sites}
Partitioning the 11 SLE cell types into disjoint panels of $\le 3$ types per site, the donor-gram
estimator recovers the full multicellular program (recovered donor-score AUC $0.958$, matching the
centralised $0.958$ and the best single site $0.954$), spanning cell types no single site observed.
The subspace correlation of $1.000$ is exact \emph{by construction} (the summed donor-Gram equals the
centralised Gram) and should be read as an algebraic identity, not an empirical estimate. Two baselines
place this. Standard fixed-feature federated PCA cannot run at all: with disjoint panels the cell types
common to every site are empty, so there is no shared feature space to align. The natural workaround,
per-site PCA followed by fusion over the shared donors, does run but only \emph{approximately} recovers
the centralised program (subspace correlation $0.977$, against $1.000$ for the donor-gram coupling),
because each site ranks components within its own panel and loses a program that is weak per panel but
strong once cell types are combined. The donor-mode coupling avoids that loss by summing the panels
before decomposition. As shown next, this capability is private at the complete-panel level provided the
donor coupling is computed under secure computation and only the loadings are released.

\subsection{Privacy: membership inference}
\begin{figure}[t]\centering
\includegraphics[width=0.8\linewidth]{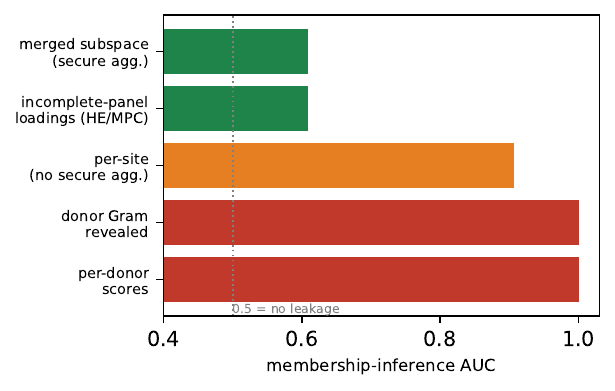}
\caption{Membership-inference AUC by release object ($0.5=$ no leakage). Secure aggregation
(merged subspace) and the loadings-only incomplete-panel release leak little ($0.61$); revealing the
donor Gram or per-donor scores is maximally leaky ($1.0$).}
\label{fig:privacy}
\end{figure}

Table~\ref{tab:mia} and Fig.~\ref{fig:privacy} report membership-inference AUC ($0.5=$ no leakage; reconstruction-residual
attack, 30 splits). For the \emph{complete-panel} scheme, a curious coordinator seeing per-site
subspaces \emph{without} secure aggregation achieves AUC $0.906$; with secure aggregation, revealing
only the merged subspace, leakage drops to $0.609$, a measurable reduction. The
\emph{incomplete-panel} scheme has two regimes. If the donor Gram is \emph{revealed}, its eigenvectors
recover the donor scores exactly (correlation $1.000$) and MIA-AUC is $1.000$, maximally leaky,
equivalent to releasing per-donor scores. But the Gram is only an intermediate: if the donor-coupling
(eigendecomposition) is computed under secure computation and \emph{only the program loadings are
released} (never the Gram or donor scores), leakage drops to $0.609$, identical to the complete-panel
scheme (Table~\ref{tab:mia}). The incomplete-panel capability is thus private at the complete-panel
level \emph{provided} the Gram step uses homomorphic encryption or multiparty computation, which is
heavier than the additive secure aggregation the complete-panel scheme requires. The $0.609$ figure is
residual leakage under one standard attack, not a worst-case bound. In deployment terms: a consortium
using the complete-panel scheme under secure aggregation discloses only population-level loadings at low
measured membership risk, whereas one that needs the incomplete-panel capability must additionally
protect the donor coupling with HE/MPC. This release-object taxonomy (which quantities are safe to
reveal, and under what cryptography) is, to our knowledge, the first privacy characterisation of
federated multicellular tensor decomposition.

\begin{figure}[t]\centering
\includegraphics[width=0.82\linewidth]{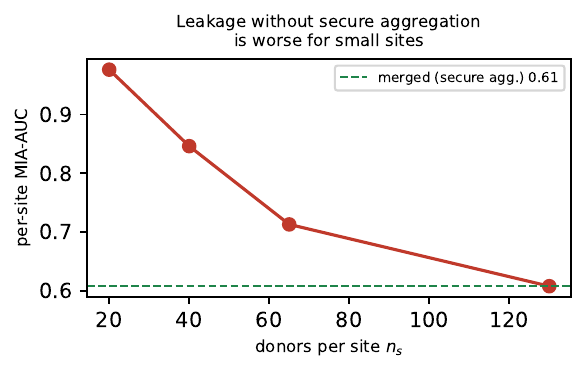}
\caption{Per-site membership leakage without secure aggregation grows as sites shrink (small local
subspaces overfit their members); secure aggregation removes this by revealing only the merged subspace
(MIA-AUC $0.61$, dashed).}
\label{fig:sitesize}
\end{figure}

Two further analyses sharpen the picture (Fig.~\ref{fig:sitesize}). First, a complementary
\emph{attribute-inference} attack: training a classifier from donor scores on the released program to
predict a sensitive attribute, the program leaks self-reported ancestry at AUC $0.81$ and sex at
$0.65$. The ancestry leakage reflects that the interferon program covaries with ancestry in this cohort,
and it is not removed by secure aggregation, so consortia should treat the released program, not only
the messages, as potentially attribute-revealing (a further argument for the DP back-end of
Section~\ref{sec:disc}). Second, the per-site (no-secure-aggregation) membership leakage \emph{grows as
sites shrink}: MIA-AUC rises from $0.61$ at $130$ donors/site to $0.98$ at $20$ donors/site, because
small local subspaces overfit their members. Secure aggregation, which prevents any party from seeing a
per-site subspace, is therefore most valuable precisely in the many-small-site federations that
motivate this work.

\subsection{When multicellularity matters: a cross-tissue map}
\label{sec:whenmatters}
The contribution is faithful private \emph{recovery}; whether the recovered program also
\emph{out-predicts} the best single cell type is a separate, tissue-dependent question, which we map
across the cohorts with a stringent test: nested donor-stratified five-fold cross-validation (the
predictive factor is selected on training donors only and scored on held-out donors), the multicellular
program benchmarked against the best single-compartment classifier selected the same way, with
donor-bootstrap $95\%$ confidence intervals on the gap.

Multicellularity wins where cross-cell-type coordination is the biology. On the interstitial-lung-disease
atlas~\cite{natri2024ild} the multicellular program attains cross-validated AUC $0.959$ against $0.910$
for the best single compartment, a gap of $+0.049$ whose $95\%$ CI $[+0.006,+0.101]$ excludes zero and
that is positive in all $10$ cross-validation seeds; the winning factor spans $\approx24$ effective cell
types, an alveolar-epithelial--macrophage--endothelial fibrotic axis that no single compartment carries.
The federated estimator recovers this program faithfully (subspace correlation $0.981$ to centralised,
AUC $0.969$), so the predictive advantage survives federation.

A second fibrotic tissue corroborates this pattern, with caveats we state plainly. On the
fibrosing-cholangiopathy liver atlas~\cite{andrews2024psc} the multicellular program attains nested
donor-stratified cross-validated AUC $0.975$ (10-seed mean), with a positive multi-minus-single gap in
all $10$ seeds; a leave-one-donor-out reanalysis gives AUC $1.000$, and its $200$-permutation
label-shuffle test gives empirical $p=0.005$ (the matched five-fold permutation gives $p=0.010$),
establishing that the separation is label-specific rather than a dimensional-overfitting artefact at
$n\ll p$. The leading factor spans $\approx23$ effective cell types,
an immune--stromal fibrotic niche (hepatic-pit, NK, and monocyte compartments top-loaded), consistent
with the lung result. We read this as corroboration, not a co-equal anchor, for four reasons. (i)~The
cohort is small ($16$ donors, only $6$ controls), below a reliable single-fold threshold. (ii)~We lead
on the multicellular separation and the permutation $p$, not on the raw multi-minus-single gap, because
the single-cell-type baseline is unstable at this $n$ ($0.56\pm0.20$ over seeds) and inflates that gap.
(iii)~A near-perfect AUC at $n=16$ cannot exclude a confounder perfectly correlated with PSC-vs-normal:
the permutation test rules out dimensional overfitting, not a real confound, though the broad,
biologically-coherent $\approx23$-cell-type program, and the separation holding for pure PSC-vs-normal
($8$ vs $6$ donors, leave-one-donor-out AUC $1.000$), argue against a narrow technical artefact or a
two-PBC-donor driver. (iv)~No
independent open replication was obtainable (candidate intestinal-failure-associated and
cystic-fibrosis-airway atlases each held only three disease cases, uninterpretable). The liver therefore
extends the lung finding from one fibrotic tissue to two, without claiming large-cohort effect sizes.

Elsewhere, where one compartment dominates, the program is at parity. The gap to the best single cell
type is $+0.013$ in SLE blood (multicellular $0.958$ vs.\ classical monocyte $0.945$), $-0.064$ in COVID
severity (a single monocyte compartment is better), $+0.008$ in Crohn's gut, $+0.024$ in kidney injury
(CI includes zero), and $-0.015$ in cardiomyopathy (cardiac fibroblast alone is better). Of seven cohorts
spanning blood, gut, kidney, heart, lung and liver, the multicellular program exceeds the best single
compartment only in the two fibrotic, strongly coordinated tissues (lung and liver). This is the honest
scope: federated multicellular recovery
earns its cost where the disease signal is genuinely distributed across cell types, and remains at
parity (still faithfully and privately recovered) where it is not.

\subsection{Robustness and communication cost}
\begin{figure}[t]\centering
\includegraphics[width=\linewidth]{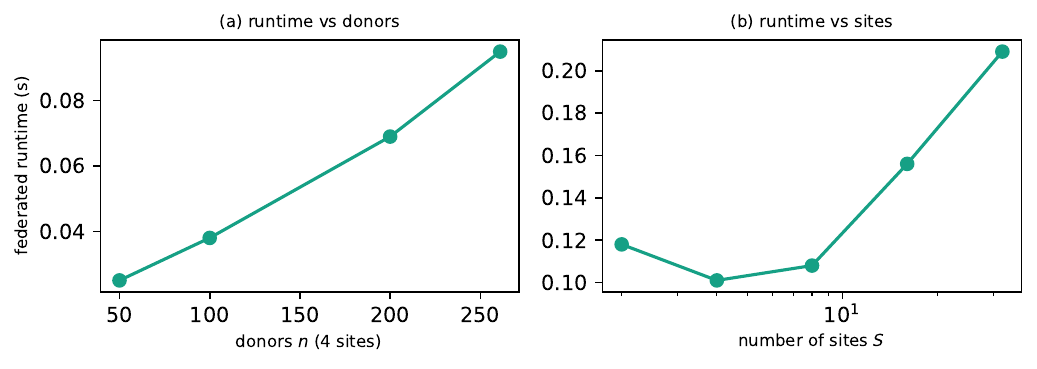}
\caption{Scalability of the federated estimator: (a) runtime grows linearly in donors and (b)
sub-linearly in the number of sites, remaining well under a second on the SLE atlas.}
\label{fig:scalability}
\end{figure}

The federated recovery is insensitive to the two main hyperparameters (Table~\ref{tab:ablation}). Across
rank $K\in\{2,\dots,20\}$ the disease-program AUC stays $0.955$--$0.958$ and ISG enrichment $\geq0.997$
(the subspace correlation fluctuates with $K$ only because lower-variance components are less stably
ordered, not the disease program). Across site counts $S\in\{2,\dots,32\}$ (down to as few as eight
donors per site), the federated AUC remains $0.957$--$0.958$ and subspace correlation $\geq0.96$,
confirming that the stacked-SVD merge tolerates many small, rank-deficient sites. On communication, each
site transmits $K\,C\,G$ floats per round ($165{,}000$ for SLE, $270{,}000$ for COVID, $135{,}000$ for
IBD at $K{=}10$), independent of its cell count and orders of magnitude below the $\sim\!10^6$ cells
$\times$ genes that pooling would transfer; the complete-panel scheme converges in a single round. End
to end the federated estimator is also computationally light (on the SLE atlas the per-site SVDs plus
merge complete in $0.10$\,s versus $0.16$\,s for the centralised SVD), so the privacy and governance
benefits carry no runtime penalty. The global-mean centering is also what makes the merge robust
to site--label confounding, the realistic failure mode of cross-institution federation. Sweeping the
case-fraction skew between two sites from balance to extreme ($10$ seeds; Table~\ref{tab:centering}),
federated global-mean centering holds the program at AUC $0.957$--$0.958$ throughout, whereas naive
per-site centering degrades monotonically, from $0.957$ at balance to $0.861$ when one site is $90\%$
cases. The merged \emph{subspace} stays similar either way (canonical correlation $\geq0.98$); what naive
centering corrupts is the disease program's donor scores, which each site shifts by a label-correlated
mean. Runtime scales gracefully (Fig.~\ref{fig:scalability}):
linearly in donors ($0.025$\,s at $n{=}50$ to $0.095$\,s at $n{=}261$) and sub-linearly in sites
($0.10$--$0.21$\,s for $S=2$ to $32$).

\begin{table}[t]
\caption{Confounding robustness (SLE, 2 sites, 10 seeds): federated phenotype-AUC vs.\ case-fraction
skew between sites, under federated global-mean vs.\ naive per-site centering. Global-mean centering is
flat; naive centering degrades with skew.}
\label{tab:centering}
\centering
\begin{tabular}{lcccccc}
\toprule
site-0 case fraction & 0.5 & 0.6 & 0.7 & 0.8 & 0.9 & 0.97 \\
\midrule
global-mean (ours) & 0.958 & 0.957 & 0.957 & 0.957 & 0.957 & 0.957 \\
naive per-site & 0.957 & 0.956 & 0.949 & 0.929 & 0.861 & 0.861 \\
\bottomrule
\end{tabular}
\end{table}

Finally, the choice of a donor-mode SVD (Tucker-type) decomposition matters: replacing it with a
trilinear CP-ALS of the same rank collapses recovery to SLE-AUC $0.534$ (best-of-$K$ $0.592$) versus
$0.958$ for the SVD. CP forces a single rank-one gene pattern shared across cell types, which underfits
the donor axis; the donor-mode SVD, which lets each program carry a full $C\times G$ loading, is
essential and is what scITD also uses.

\begin{table}[t]
\caption{Ablations (SLE). Recovery is robust to rank $K$ and to the number of sites $S$ (down to 8
donors/site). Centralised AUC $=0.958$.}
\label{tab:ablation}
\centering
\begin{tabular}{cccc|ccc}
\toprule
\multicolumn{4}{c|}{rank $K$ (4 sites)} & \multicolumn{3}{c}{sites $S$ ($K{=}10$)} \\
$K$ & subcorr & fed AUC & ISG & $S$ & subcorr & fed AUC \\
\midrule
2  & 0.989 & 0.955 & 0.997 & 2  & 0.979 & 0.958 \\
5  & 0.935 & 0.956 & 0.998 & 4  & 0.963 & 0.957 \\
8  & 0.979 & 0.958 & 0.998 & 8  & 0.970 & 0.957 \\
10 & 0.967 & 0.958 & 0.999 & 16 & 0.989 & 0.957 \\
20 & 0.968 & 0.958 & 0.998 & 32 & 1.000 & 0.958 \\
\bottomrule
\end{tabular}
\end{table}

\subsection{Cross-disease federated meta-analysis}
\begin{figure}[t]\centering
\includegraphics[width=0.82\linewidth]{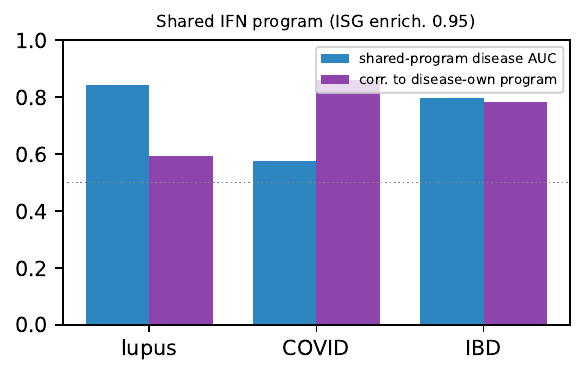}
\caption{Cross-disease federated meta-analysis. A shared interferon-type multicellular program
(ISG enrichment $0.95$), recovered federated across lupus, COVID and IBD as three disease-sites,
associates with disease (blue) and aligns with each disease's own program (purple).}
\label{fig:crossdisease}
\end{figure}

A practical use of the framework is to identify programs \emph{shared across diseases} held at
different institutions. We treat the three atlases as three disease-sites, harmonise them to a common
space of four immune compartments (B, T, monocyte, plasma; present in all annotations) and $800$
top-variable genes from the $22{,}644$ shared by all three, and federate (donors are fully disjoint
across diseases). The merge recovers a single \emph{shared multicellular program} that is strongly
interferon-typed (ISG enrichment AUC $0.948$) and associates with disease in lupus (AUC $0.842$) and
Crohn's inflammation (AUC $0.796$), more weakly in COVID severity (AUC $0.576$); its gene-mode loading
aligns with each disease's own program (correlation $0.59$/$0.86$/$0.78$ for lupus/COVID/IBD;
Fig.~\ref{fig:crossdisease}). A shared interferon-type immune axis is thus recoverable across
autoimmune, viral, and inflammatory disease without pooling cells, an analysis impossible to run
centrally on data that cannot be co-located. We report this as a feasibility demonstration; the weak
COVID association indicates the shared axis captures interferon biology common to the three rather than
COVID-specific severity drivers.

\begin{table}[t]
\caption{Federated vs.\ centralised recovery (SLE case/control AUC; COVID severity program). Centralised
SLE-AUC $=0.958$. The COVID figure is the in-sample best-program AUC; the cross-validated severity AUC
is $0.678$ (Sec.~III).}
\label{tab:rec}
\centering
\begin{tabular}{lcccc}
\toprule
Partition & Sites & Fed.\ AUC & Subspace corr. & ISG enr. \\
\midrule
SLE, IID & 4 & 0.958 & 0.967 & 0.999 \\
SLE, IID & 8 & 0.957 & 0.974 & 0.998 \\
SLE, multi-ancestry & 2 & 0.958 & 0.981 & 0.998 \\
SLE, processing cohort & 4 & 0.956 & 0.980 & 0.998 \\
COVID, real sites & 3 & 0.700 & 0.989 & n/a \\
\bottomrule
\end{tabular}
\end{table}

\begin{table}[t]
\caption{Membership-inference attack AUC by release object ($0.5=$ no leakage; reconstruction-residual
attack, 30 splits, SLE).}
\label{tab:mia}
\centering
\begin{tabular}{lc}
\toprule
Released object & MIA-AUC (95\% CI) \\
\midrule
Complete-panel: per-site subspaces (no secure agg.) & 0.906 [0.84, 0.96] \\
Complete-panel: merged subspace (with secure agg.) & \textbf{0.609 [0.55, 0.68]} \\
Incomplete-panel: loadings only (Gram under HE/MPC) & \textbf{0.609} \\
Incomplete-panel: donor Gram revealed & 1.000 \\
\textit{(reference)} per-donor scores & 1.000 \\
\bottomrule
\end{tabular}
\end{table}

\begin{figure}[t]\centering
\includegraphics[width=\linewidth]{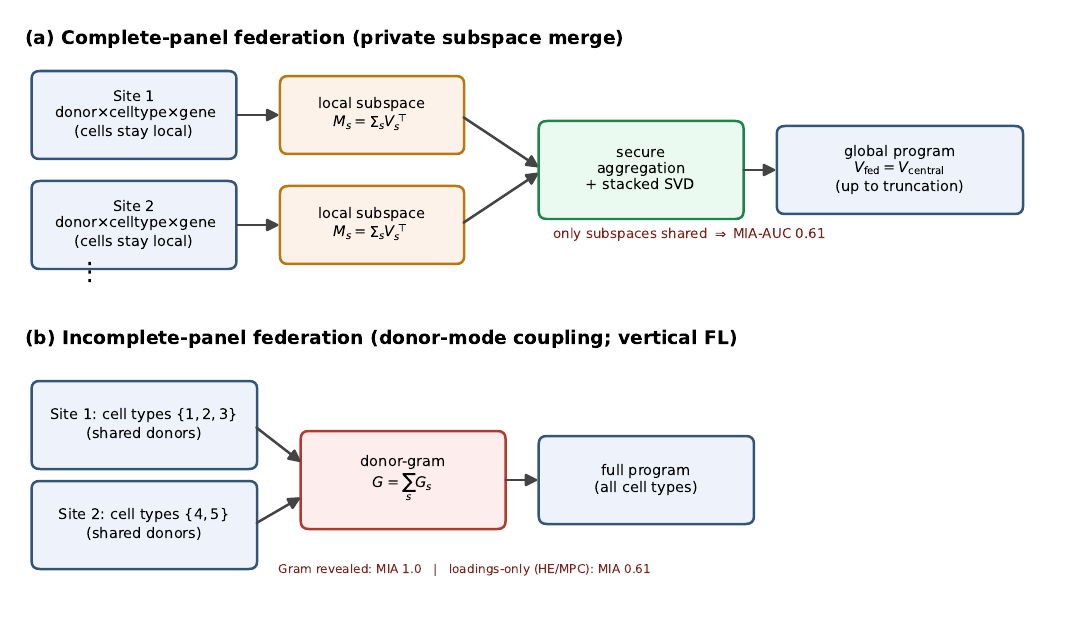}
\caption{Federated multicellular tensor decomposition. (a) Complete-panel: sites share only local
program subspaces; the secure-aggregation stacked-SVD merge reproduces the centralised program
(membership-inference AUC $0.61$). (b) Incomplete-panel (vertical FL): donor-mode coupling recovers
programs spanning cell types no single site observed. Revealing the donor Gram exposes donor scores
(membership-inference AUC $1.0$); computing the coupling under HE/MPC and releasing only the loadings
restores complete-panel privacy (AUC $0.61$).}
\label{fig:1}
\end{figure}

\section{Discussion}
\label{sec:disc}
We have shown that multicellular immune programs can be recovered by a federated tensor decomposition
that shares only low-dimensional program subspaces, is compatible with secure aggregation, returns the
centralised result up to truncation, and additionally recovers programs from cell-type-incomplete
sites. A membership-inference evaluation demonstrates the privacy benefit of secure aggregation, and a
feasibility analysis delineates where differential privacy becomes possible.

\emph{Practical implications.} Multicellular-program analyses such as scITD have proven informative on
centralised atlases but are blocked, by data-protection constraints, from the multi-institution and
multi-ancestry cohorts where they would be most powerful. Our results show those analyses can be run
faithfully across such cohorts without moving cells: a consortium can agree a gene panel and rank,
exchange only program subspaces under secure aggregation, and obtain the same programs a pooled
analysis would yield, with quantified membership leakage (low for the complete-panel scheme). The
multi-ancestry and real three-site experiments indicate this is robust to the label skew and small,
heterogeneous sites that real federations exhibit.

Three limitations bound the contribution. \emph{First, the predictive value of multicellularity is
tissue-dependent.} On compartment-dominated blood, gut, kidney and heart cohorts the dominant disease
program is also legible from a single cell type (e.g.\ classical monocytes give AUC $0.945$ for SLE;
cardiac fibroblast alone exceeds the joint factor in cardiomyopathy), so there the contribution is
faithful private federated \emph{recovery} of established programs and the incomplete-panel capability,
not added prediction. Where cross-cell-type coordination is strong, multicellularity \emph{does} add
predictive power (interstitial lung disease, \S\ref{sec:whenmatters}), confirming the method earns its
cost when the disease signal is genuinely distributed; whether that regime is the common or the
exceptional case across diseases remains open, demonstrated here in two strongly coordinated fibrotic
tissues (lung and liver). The liver is corroborating rather than co-equal: it rests on a single small
cohort ($n=16$, $6$ controls) without an independent open replication, and at that $n$ a near-perfect
separation cannot fully exclude a confounder (the permutation test excludes overfitting, not
confounding), so the well-powered lung cohort ($n=109$) remains the anchor. \emph{Second, formal differential privacy is out of
reach at current cohort sizes.} Donor-level DP~\cite{dwork2014algorithmic,dwork2006calibrating} on the released program requires $n=\widetilde O(d)$
samples in the ambient program dimension~\cite{noisypower,dppca2022,singhal2021privately}, whereas single-cell atlases have
$n\approx10^2$; on these cohorts the noisy power method fails to recover the program at any meaningful
budget, and the apparent escape of restricting the private step to a known public gene signature
succeeds only because that program is already public, not as a private-recovery mechanism. Secure
aggregation therefore provides the operative (lossless) guarantee for the complete-panel scheme, with
its residual leakage measured by the membership-inference attack, and formal DP, with privacy accounted under zero-concentrated DP composition~\cite{bun2016concentrated}, becomes realistic only
at consortium scale ($n\gtrsim10^3$). \emph{Third, the
incomplete-panel capability requires stronger cryptography}: revealing the donor Gram exposes donor
scores (MIA-AUC $1.000$), but computing the coupling under homomorphic encryption / multiparty
computation and releasing only the program loadings restores complete-panel privacy (MIA-AUC $0.609$),
heavier than additive secure aggregation; it also presumes donors are linked across sites (a
private-set-intersection problem) under harmonised cell-type annotations.
Further simplifications: secure aggregation is modelled rather than deployed cryptographically, and we
report no end-to-end systems experiment (communication is $K\times CG$ floats/round; dropout-robust
secure aggregation is assumed).

\emph{Future work.} The negative control points to where multicellularity should matter most: settings
with genuine cross-cell-type coordination rather than one dominant compartment. Two are natural.
(i)~\emph{Multi-omic coupling}, in which paired modalities (e.g.\ RNA and chromatin accessibility, or surface
protein) couple irreducibly through shared regulatory structure, and the vertical-FL form of our
incomplete-panel scheme (parties holding different modalities of shared donors) applies directly, with
the loadings-only release giving complete-panel privacy under HE/MPC. (ii)~\emph{Spatial niches}, where
a coordinated multi-cell-type signature is the object of interest by construction. Methodologically,
adding a differentially private or homomorphic-encryption back-end to the donor-Gram coupling~\cite{abadi2016deep} (so the
incomplete-panel capability attains formal, not merely empirical, privacy) and developing
confounding-robust aggregation for severe site--label skew are the main open directions. The
multicellular-tensor view is also complementary to multi-modal matrix factorisation
(e.g.\ MOFA~\cite{mofa}) and cell--cell communication tensors~\cite{tensorcell2cell}, whose federation
we leave to future work.

\section{Conclusion}
Federated multicellular tensor decomposition recovers established immune programs across simulated and
real institutional boundaries and across ancestries, recovers programs from sites with incomplete
cell-type panels, and does so while sharing only low-dimensional subspaces under secure aggregation
with quantified privacy. It is a practical step toward privacy-preserving, multi-cohort single-cell
immunology.

\section*{Data Availability}
Public via CZ CELLxGENE: SLE collection \texttt{436154da}; COVID-19 collection \texttt{ddfad306};
IBDverse collection \texttt{7f7fdf50}; PSC/PBC liver collection \texttt{0c8a364b}.
\section*{Code Availability}
Pipeline at \todo{REPO URL}.
\section*{Ethics}
Reuse of publicly available de-identified data under the original studies' approvals; no new human data.

\bibliographystyle{IEEEtran}
\bibliography{refs}

@article{scitd,
  author  = {Mitchel, Jonathan and Gordon, M. Grace and Perez, Richard K. and Biederstedt, Evan and Bueno, Raymund and Ye, Chun Jimmie and Kharchenko, Peter V.},
  title   = {Coordinated, multicellular patterns of transcriptional variation that stratify patient cohorts are revealed by tensor decomposition},
  journal = {Nature Biotechnology},
  volume  = {43},
  pages   = {1192--1201},
  year    = {2025},
  doi     = {10.1038/s41587-024-02411-z}
}

@article{perez2022,
  author  = {Perez, Richard K. and Gordon, M. Grace and Subramaniam, Meena and Kim, Min Cheol and Hartoularos, George C. and Targ, Sasha and Sun, Yang and Ogorodnikov, Anton and Bueno, Raymund and others},
  title   = {Single-cell RNA-seq reveals cell type--specific molecular and genetic associations to lupus},
  journal = {Science},
  volume  = {376},
  number  = {6589},
  pages   = {eabf1970},
  year    = {2022},
  doi     = {10.1126/science.abf1970}
}

@article{stephenson2021,
  author  = {Stephenson, Emily and Reynolds, Gary and Botting, Rachel A. and others},
  title   = {Single-cell multi-omics analysis of the immune response in COVID-19},
  journal = {Nature Medicine},
  volume  = {27},
  pages   = {904--916},
  year    = {2021},
  doi     = {10.1038/s41591-021-01329-2}
}

@inproceedings{secagg,
  author    = {Bonawitz, Keith and Ivanov, Vladimir and Kreuter, Ben and Marcedone, Antonio and McMahan, H. Brendan and Patel, Sarvar and Ramage, Daniel and Segal, Aaron and Seth, Karn},
  title     = {Practical Secure Aggregation for Privacy-Preserving Machine Learning},
  booktitle = {Proc.\ 2017 ACM SIGSAC Conf.\ on Computer and Communications Security (CCS)},
  pages     = {1175--1191},
  year      = {2017},
  doi       = {10.1145/3133956.3133982}
}

@inproceedings{noisypower,
  author    = {Hardt, Moritz and Price, Eric},
  title     = {The Noisy Power Method: A Meta Algorithm with Applications},
  booktitle = {Advances in Neural Information Processing Systems (NeurIPS)},
  volume    = {27},
  pages     = {2861--2869},
  year      = {2014}
}

@article{tensorcell2cell,
  author  = {Armingol, Erick and Baghdassarian, Hratch M. and Martino, Cameron and Perez-Lopez, Araceli and Aamodt, Caitlin and Knight, Rob and Lewis, Nathan E.},
  title   = {Context-aware deconvolution of cell--cell communication with Tensor-cell2cell},
  journal = {Nature Communications},
  volume  = {13},
  pages   = {3665},
  year    = {2022},
  doi     = {10.1038/s41467-022-31369-2}
}

@article{parafac2rise,
  author  = {Chen, Andrew and others},
  title   = {Integrative, high-resolution analysis of single-cell gene expression across experimental conditions with PARAFAC2-RISE},
  journal = {Cell Systems},
  year    = {2025},
  doi     = {10.1016/j.cels.2025.101312},
  note    = {Complete author list before submission}
}

@article{fedscgen,
  author  = {Bakhtiari, Mohammad and Bonn, Stefan and Theis, Fabian and Zolotareva, Olga and Baumbach, Jan},
  title   = {FedscGen: privacy-preserving federated batch effect correction of single-cell RNA sequencing data},
  journal = {Genome Biology},
  year    = {2025},
  doi     = {10.1186/s13059-025-03684-6}
}

@article{pricell,
  author  = {Sav, Sinem and Bossuat, Jean-Philippe and Troncoso-Pastoriza, Juan R. and Claassen, Manfred and Hubaux, Jean-Pierre},
  title   = {Privacy-preserving federated neural network learning for disease-associated cell classification},
  journal = {Patterns},
  volume  = {3},
  number  = {5},
  pages   = {100487},
  year    = {2022},
  doi     = {10.1016/j.patter.2022.100487}
}

@article{tabula,
  author  = {Wang, Jiayuan and others},
  title   = {Toward a privacy-preserving predictive foundation model of single-cell transcriptomics with federated learning and tabular modeling},
  journal = {bioRxiv},
  year    = {2025},
  doi     = {10.1101/2025.01.06.631427},
  note    = {Preprint; complete/verify author list before submission}
}

@article{distpca,
  author  = {Fan, Jianqing and Wang, Dong and Wang, Kaizheng and Zhu, Ziwei},
  title   = {Distributed estimation of principal eigenspaces},
  journal = {Annals of Statistics},
  volume  = {47},
  number  = {6},
  pages   = {3009--3031},
  year    = {2019},
  doi     = {10.1214/18-AOS1713}
}

@article{privateqtl,
  author  = {others},
  title   = {Secure and federated quantitative trait loci mapping with privateQTL},
  journal = {Cell Genomics},
  year    = {2025},
  note    = {Real (PMID 39947138, Cell Genomics 2025, PII S2666-979X(25)00025-4); complete author list + exact DOI at submission}
}

@article{ibdverse,
  author  = {others},
  title   = {Cell-type-resolved genetic variation shapes inflammatory bowel disease risk},
  journal = {Nature},
  year    = {2026},
  doi     = {10.1038/s41586-026-10627-z},
  note    = {IBDverse atlas (Wellcome Sanger); complete author list at submission}
}

@article{mofa,
  author  = {Argelaguet, Ricard and Arnol, Damien and Bredikhin, Danila and Deloro, Yonatan and Velten, Britta and Marioni, John C. and Stegle, Oliver},
  title   = {MOFA+: a statistical framework for comprehensive integration of multi-modal single-cell data},
  journal = {Genome Biology},
  volume  = {21},
  pages   = {111},
  year    = {2020},
  doi     = {10.1186/s13059-020-02015-1}
}

@article{walker2024leakage,
  author  = {Walker, Conor R. and Li, Xiaoting and Chakravarthy, Manav and Lounsbery-Scaife, William and Choi, Yoolim A. and Singh, Ritambhara and G\"ursoy, Gamze},
  title   = {Private information leakage from single-cell count matrices},
  journal = {Cell},
  year    = {2024},
  note    = {PMID 39362221; DOI 10.1016/j.cell.2024.09.012}
}

@inproceedings{dppca2022,
  author    = {Liu, Xiyang and Kong, Weihao and Jain, Prateek and Oh, Sewoong},
  title     = {{DP-PCA}: Statistically Optimal and Differentially Private {PCA}},
  booktitle = {Advances in Neural Information Processing Systems (NeurIPS)},
  year      = {2022},
  note      = {arXiv:2205.13709}
}

@article{natri2024ild,
  author  = {Natri, Heini M. and Del Azodi, Christina B. and Peter, Lance and Taylor, Chase J. and Chugh, Sagrika and Kendle, Robert and Chung, Mei-i and Flaherty, David K. and Matlock, Brittany K. and Calvi, Carla L. and Blackwell, Timothy S. and Ware, Lorraine B. and Bacchetta, Matthew and Walia, Rajat and Shaver, Ciara M. and Kropski, Jonathan A. and McCarthy, Davis J. and Banovich, Nicholas E.},
  title   = {Cell-type-specific and disease-associated expression quantitative trait loci in the human lung},
  journal = {Nature Genetics},
  year    = {2024},
  note    = {DOI 10.1038/s41588-024-01702-0; GEO GSE227136}
}

@article{andrews2024psc,
  author  = {Andrews, Tallulah S. and Nakib, Diana and Perciani, Catia T. and Ma, Xue Zhong and Liu, Lewis and Winter, Erin and Camat, Damra and Chung, Sai W. and Lumanto, Patricia and Manuel, Justin and Mangroo, Shantel and Hansen, Bettina and Arpinder, Bal and Thoeni, Cornelia and Sayed, Blayne and Feld, Jordan and Gehring, Adam and Gulamhusein, Aliya and Hirschfield, Gideon M. and Ricciuto, Amanda and Bader, Gary D. and McGilvray, Ian D. and MacParland, Sonya},
  title   = {Single-cell, single-nucleus, and spatial transcriptomics characterization of the immunological landscape in the healthy and PSC human liver},
  journal = {Journal of Hepatology},
  year    = {2024},
  doi     = {10.1016/j.jhep.2023.12.023},
  note    = {CZ CELLxGENE collection 0c8a364b; GEO GSE243977}
}

@inproceedings{mcmahan2017fedavg,
  author    = {McMahan, H. Brendan and Moore, Eider and Ramage, Daniel and Hampson, Seth and Ag\"uera y Arcas, Blaise},
  title     = {Communication-Efficient Learning of Deep Networks from Decentralized Data},
  booktitle = {Proc.\ 20th Int.\ Conf.\ on Artificial Intelligence and Statistics (AISTATS), PMLR 54},
  pages     = {1273--1282},
  year      = {2017},
  url       = {https://proceedings.mlr.press/v54/mcmahan17a.html}
}

@article{kairouz2021advances,
  author  = {Kairouz, Peter and McMahan, H. Brendan and Avent, Brendan and Bellet, Aur\'elien and others},
  title   = {Advances and Open Problems in Federated Learning},
  journal = {Foundations and Trends in Machine Learning},
  volume  = {14},
  number  = {1--2},
  pages   = {1--210},
  year    = {2021},
  doi     = {10.1561/2200000083}
}

@article{yang2019federated,
  author  = {Yang, Qiang and Liu, Yang and Chen, Tianjian and Tong, Yongxin},
  title   = {Federated Machine Learning: Concept and Applications},
  journal = {ACM Transactions on Intelligent Systems and Technology},
  volume  = {10},
  number  = {2},
  pages   = {12:1--12:19},
  year    = {2019},
  doi     = {10.1145/3298981}
}

@article{rieke2020future,
  author  = {Rieke, Nicola and Hancox, Jonny and Li, Wenqi and Milletar\`i, Fausto and Roth, Holger R. and Albarqouni, Shadi and Bakas, Spyridon and Galtier, Mathieu N. and Landman, Bennett A. and Maier-Hein, Klaus and Ourselin, S\'ebastien and Sheller, Micah and Summers, Ronald M. and Trask, Andrew and Xu, Daguang and Baust, Maximilian and Cardoso, M. Jorge},
  title   = {The future of digital health with federated learning},
  journal = {npj Digital Medicine},
  volume  = {3},
  pages   = {119},
  year    = {2020},
  doi     = {10.1038/s41746-020-00323-1}
}

@article{sheller2020federated,
  author  = {Sheller, Micah J. and Edwards, Brandon and Reina, G. Anthony and Martin, Jason and Pati, Sarthak and Kotrotsou, Aikaterini and Milchenko, Mikhail and Xu, Weilin and Marcus, Daniel and Colen, Rivka R. and Bakas, Spyridon},
  title   = {Federated learning in medicine: facilitating multi-institutional collaborations without sharing patient data},
  journal = {Scientific Reports},
  volume  = {10},
  pages   = {12598},
  year    = {2020},
  doi     = {10.1038/s41598-020-69250-1}
}

@article{dwork2014algorithmic,
  author  = {Dwork, Cynthia and Roth, Aaron},
  title   = {The Algorithmic Foundations of Differential Privacy},
  journal = {Foundations and Trends in Theoretical Computer Science},
  volume  = {9},
  number  = {3--4},
  pages   = {211--487},
  year    = {2014},
  doi     = {10.1561/0400000042}
}

@inproceedings{dwork2006calibrating,
  author    = {Dwork, Cynthia and McSherry, Frank and Nissim, Kobbi and Smith, Adam},
  title     = {Calibrating Noise to Sensitivity in Private Data Analysis},
  booktitle = {Theory of Cryptography (TCC), LNCS 3876},
  pages     = {265--284},
  year      = {2006},
  doi       = {10.1007/11681878_14}
}

@inproceedings{dwork2014analyzegauss,
  author    = {Dwork, Cynthia and Talwar, Kunal and Thakurta, Abhradeep and Zhang, Li},
  title     = {Analyze {Gauss}: Optimal Bounds for Privacy-Preserving Principal Component Analysis},
  booktitle = {Proc.\ 46th Annual ACM Symposium on Theory of Computing (STOC)},
  pages     = {11--20},
  year      = {2014},
  doi       = {10.1145/2591796.2591883}
}

@inproceedings{abadi2016deep,
  author    = {Abadi, Martin and Chu, Andy and Goodfellow, Ian and McMahan, H. Brendan and Mironov, Ilya and Talwar, Kunal and Zhang, Li},
  title     = {Deep Learning with Differential Privacy},
  booktitle = {Proc.\ 2016 ACM SIGSAC Conf.\ on Computer and Communications Security (CCS)},
  pages     = {308--318},
  year      = {2016},
  doi       = {10.1145/2976749.2978318}
}

@article{chaudhuri2013near,
  author  = {Chaudhuri, Kamalika and Sarwate, Anand D. and Sinha, Kaushik},
  title   = {A Near-Optimal Algorithm for Differentially-Private Principal Components},
  journal = {Journal of Machine Learning Research},
  volume  = {14},
  pages   = {2905--2943},
  year    = {2013},
  url     = {https://jmlr.org/papers/v14/chaudhuri13a.html}
}

@inproceedings{bun2016concentrated,
  author    = {Bun, Mark and Steinke, Thomas},
  title     = {Concentrated Differential Privacy: Simplifications, Extensions, and Lower Bounds},
  booktitle = {Theory of Cryptography (TCC-B), LNCS 9985},
  pages     = {635--658},
  year      = {2016},
  doi       = {10.1007/978-3-662-53641-4_24}
}

@inproceedings{singhal2021privately,
  author    = {Singhal, Vikrant and Steinke, Thomas},
  title     = {Privately Learning Subspaces},
  booktitle = {Advances in Neural Information Processing Systems (NeurIPS) 34},
  year      = {2021},
  url       = {https://proceedings.neurips.cc/paper/2021/hash/09b69adcd7cbae914c6204984097d2da-Abstract.html}
}

@inproceedings{shokri2017membership,
  author    = {Shokri, Reza and Stronati, Marco and Song, Congzheng and Shmatikov, Vitaly},
  title     = {Membership Inference Attacks Against Machine Learning Models},
  booktitle = {Proc.\ 2017 IEEE Symposium on Security and Privacy (S\&P)},
  pages     = {3--18},
  year      = {2017},
  doi       = {10.1109/SP.2017.41}
}

@inproceedings{carlini2022membership,
  author    = {Carlini, Nicholas and Chien, Steve and Nasr, Milad and Song, Shuang and Terzis, Andreas and Tram\`er, Florian},
  title     = {Membership Inference Attacks From First Principles},
  booktitle = {Proc.\ 2022 IEEE Symposium on Security and Privacy (S\&P)},
  pages     = {1897--1914},
  year      = {2022},
  doi       = {10.1109/SP46214.2022.9833649}
}

@article{homer2008resolving,
  author  = {Homer, Nils and Szelinger, Szabolcs and Redman, Margot and Duggan, David and Tembe, Waibhav and Muehling, Jill and Pearson, John V. and Stephan, Dietrich A. and Nelson, Stanley F. and Craig, David W.},
  title   = {Resolving Individuals Contributing Trace Amounts of {DNA} to Highly Complex Mixtures Using High-Density {SNP} Genotyping Microarrays},
  journal = {PLoS Genetics},
  volume  = {4},
  number  = {8},
  pages   = {e1000167},
  year    = {2008},
  doi     = {10.1371/journal.pgen.1000167}
}

@article{erlich2014routes,
  author  = {Erlich, Yaniv and Narayanan, Arvind},
  title   = {Routes for breaching and protecting genetic privacy},
  journal = {Nature Reviews Genetics},
  volume  = {15},
  number  = {6},
  pages   = {409--421},
  year    = {2014},
  doi     = {10.1038/nrg3723}
}

@inproceedings{gentry2009fully,
  author    = {Gentry, Craig},
  title     = {Fully Homomorphic Encryption Using Ideal Lattices},
  booktitle = {Proc.\ 41st Annual ACM Symposium on Theory of Computing (STOC)},
  pages     = {169--178},
  year      = {2009},
  doi       = {10.1145/1536414.1536440}
}

@inproceedings{cheon2017homomorphic,
  author    = {Cheon, Jung Hee and Kim, Andrey and Kim, Miran and Song, Yongsoo},
  title     = {Homomorphic Encryption for Arithmetic of Approximate Numbers},
  booktitle = {Advances in Cryptology -- ASIACRYPT 2017, Part I, LNCS 10624},
  pages     = {409--437},
  year      = {2017},
  doi       = {10.1007/978-3-319-70694-8_15}
}

@inproceedings{mohassel2017secureml,
  author    = {Mohassel, Payman and Zhang, Yupeng},
  title     = {{SecureML}: A System for Scalable Privacy-Preserving Machine Learning},
  booktitle = {Proc.\ 2017 IEEE Symposium on Security and Privacy (S\&P)},
  pages     = {19--38},
  year      = {2017},
  doi       = {10.1109/SP.2017.12}
}

@article{kolda2009tensor,
  author  = {Kolda, Tamara G. and Bader, Brett W.},
  title   = {Tensor Decompositions and Applications},
  journal = {SIAM Review},
  volume  = {51},
  number  = {3},
  pages   = {455--500},
  year    = {2009},
  doi     = {10.1137/07070111X}
}

@article{delathauwer2000multilinear,
  author  = {De Lathauwer, Lieven and De Moor, Bart and Vandewalle, Joos},
  title   = {A Multilinear Singular Value Decomposition},
  journal = {SIAM Journal on Matrix Analysis and Applications},
  volume  = {21},
  number  = {4},
  pages   = {1253--1278},
  year    = {2000},
  doi     = {10.1137/S0895479896305696}
}

@article{jerbyarnon2022dialogue,
  author  = {Jerby-Arnon, Livnat and Regev, Aviv},
  title   = {{DIALOGUE} maps multicellular programs in tissue from single-cell or spatial transcriptomics data},
  journal = {Nature Biotechnology},
  volume  = {40},
  pages   = {1467--1477},
  year    = {2022},
  doi     = {10.1038/s41587-022-01288-0}
}

@article{wolf2018scanpy,
  author  = {Wolf, F. Alexander and Angerer, Philipp and Theis, Fabian J.},
  title   = {{SCANPY}: large-scale single-cell gene expression data analysis},
  journal = {Genome Biology},
  volume  = {19},
  pages   = {15},
  year    = {2018},
  doi     = {10.1186/s13059-017-1382-0}
}

@article{hao2021integrated,
  author  = {Hao, Yuhan and Hao, Stephanie and Andersen-Nissen, Erica and Mauck III, William M. and Zheng, Shiwei and Butler, Andrew and Lee, Maddie J. and Wilk, Aaron J. and Darby, Charlotte and Zager, Michael and others},
  title   = {Integrated analysis of multimodal single-cell data},
  journal = {Cell},
  volume  = {184},
  number  = {13},
  pages   = {3573--3587},
  year    = {2021},
  doi     = {10.1016/j.cell.2021.04.048}
}

@article{regev2017human,
  author  = {Regev, Aviv and Teichmann, Sarah A. and Lander, Eric S. and Amit, Ido and Benoist, Christophe and Birney, Ewan and Bodenmiller, Bernd and Campbell, Peter and Carninci, Piero and Clatworthy, Menna and others},
  title   = {The Human Cell Atlas},
  journal = {eLife},
  volume  = {6},
  pages   = {e27041},
  year    = {2017},
  doi     = {10.7554/eLife.27041}
}

@article{crowell2020muscat,
  author  = {Crowell, Helena L. and Soneson, Charlotte and Germain, Pierre-Luc and Calini, Daniela and Collin, Ludovic and Raposo, Catarina and Malhotra, Dheeraj and Robinson, Mark D.},
  title   = {muscat detects subpopulation-specific state transitions from multi-sample multi-condition single-cell transcriptomics data},
  journal = {Nature Communications},
  volume  = {11},
  pages   = {6077},
  year    = {2020},
  doi     = {10.1038/s41467-020-19894-4}
}

@article{squair2021confronting,
  author  = {Squair, Jordan W. and Gautier, Matthieu and Kathe, Claudia and Anderson, Mark A. and James, Nicholas D. and Hutson, Thomas H. and Hudelle, R\'emi and Qaiser, Taha and Matson, Kaya J. E. and Barraud, Quentin and Levine, Ariel J. and La Manno, Gioele and Skinnider, Michael A. and Courtine, Gr\'egoire},
  title   = {Confronting false discoveries in single-cell differential expression},
  journal = {Nature Communications},
  volume  = {12},
  pages   = {5692},
  year    = {2021},
  doi     = {10.1038/s41467-021-25960-2}
}

@inproceedings{grammenos2020federated,
  author    = {Grammenos, Andreas and Mendoza-Smith, Rodrigo and Crowcroft, Jon and Mascolo, Cecilia},
  title     = {Federated Principal Component Analysis},
  booktitle = {Advances in Neural Information Processing Systems (NeurIPS) 33},
  year      = {2020},
  url       = {https://proceedings.neurips.cc/paper/2020/hash/47a658229eb2368a99f1d032c8848542-Abstract.html}
}
\end{document}